\documentclass[twocolumn,showpacs,aps,prl,superscriptaddress,floatfix]{revtex4}
\usepackage{graphicx}
\usepackage{multirow}
\usepackage{dcolumn}
\usepackage{bm}
\usepackage{amsmath}
\usepackage{units}

\newcommand{\BABARPubYear}    {07}
\newcommand{\BABARPubNumber}  {038}
\newcommand{\SLACPubNumber} {12676}
\newcommand{\LANLNumber} {0707.2980}

\input pubboard/babarsym

\newcommand{\Btag}{\ensuremath{\B_\mathrm{tag}}}
\newcommand{\Brec}{\ensuremath{\B_{\CP}}}
\newcommand{\Bztojpsiks}{\ensuremath{\Bz\to\jpsi\KS}}

\newcommand{\sintwobeta} {\ensuremath{\sin 2\beta}}

\newcommand{\Bztokspiz} {\ensuremath{\Bz \to \KS\piz}}
\newcommand{\Bztokzpiz} {\ensuremath{\Bz \to \Kz\piz}}
\newcommand{\ckspiz} {\ensuremath{C_{\KS\piz}}} \newcommand{\skspiz}
{\ensuremath{S_{\KS\piz}}} \def\cf {\ensuremath{C_f}} \def\sf
{\ensuremath{S_f}} 
 
\newcommand{\Bflav}
{\ensuremath{B_{\rm flav}}} \newcommand{\zrec}{\ensuremath{z_{\CP}}}
\newcommand{\ztag}{\ensuremath{z_\mathrm{tag}}}
\newcommand{\mmiss}{\ensuremath{m_\text{miss}}}
\newcommand{\mb}{\ensuremath{m_{B}}}

\newcommand{\thetacms}{\ensuremath{\theta_{B}^*}}
\newcommand{\costhetacms}{\ensuremath{\cos\thetacms}}
\newcommand{\deltatrec}{\ensuremath{{\rm \Delta}t_{\rm r}}\xspace}

\long\def\inst#1{\par\nobreak\kern 4pt\nobreak
    {\it #1}\par\vskip 10pt plus 3pt minus 3pt}

\begin{document}

\begin{flushleft}
\babar-PUB-\BABARPubYear/\BABARPubNumber \\
SLAC-PUB-\SLACPubNumber \\
hep-ex/\LANLNumber \\
\end{flushleft}

\title{Measurement of the \CP-Violating Asymmetries 
  in \Bztokspiz\ and of the Branching Fraction of \Bztokzpiz\ }

%
%
\author{B.~Aubert}
\author{M.~Bona}
\author{D.~Boutigny}
\author{Y.~Karyotakis}
\author{J.~P.~Lees}
\author{V.~Poireau}
\author{X.~Prudent}
\author{V.~Tisserand}
\author{A.~Zghiche}
\affiliation{Laboratoire de Physique des Particules, IN2P3/CNRS et Universit\'e de Savoie, F-74941 Annecy-Le-Vieux, France }
\author{J.~Garra~Tico}
\author{E.~Grauges}
\affiliation{Universitat de Barcelona, Facultat de Fisica, Departament ECM, E-08028 Barcelona, Spain }
% ...
\collaboration{The \babar\ Collaboration}
\noaffiliation

\begin{abstract}
  We present a measurement of the time-dependent $C\!P$-violating
  asymmetries in $B^0 \to K^0_{\scriptscriptstyle S}\pi^0$\ decays
  based on $383$ million $\Upsilon(4S)\to B\kern 0.18em\overline{\kern
  -0.18em B}$ events collected by the \mbox{\slshape
  B\kern-0.1em{\smaller A}\kern-0.1em B\kern-0.1em{\smaller
  A\kern-0.2em R}}\ experiment at the PEP-II asymmetric-energy $B$
  Factory at SLAC. We measure the direct $C\!P$-violating asymmetry
  $C_{K^0_{\scriptscriptstyle S}\pi^0} = 0.24 \pm 0.15 \pm 0.03$ and
  the CP-violating asymmetry in the interference between mixing and
  decay $S_{K^0_{\scriptscriptstyle S}\pi^0} = 0.40 \pm 0.23 \pm 0.03$,
  where the first errors are statistical and the second are systematic. On
  the same sample, we measure the decay branching fraction, obtaining
  ${\cal B}(B^0 \to K^0\pi^0)= (10.3 \pm 0.7
  \pm 0.6)\times 10^{-6}$. 

\end{abstract}


\maketitle


The \babar{} and Belle experiments have measured the weak phase
$\beta$~\cite{BaBarSin2betaObs,BelleSin2betaObs} of the
Cabibbo-Kobayashi-Maskawa (CKM) quark mixing matrix~\cite{CKM} with a
better precision than the Standard Model (SM) prediction~\cite{fits}
derived from measurements of other CP-conserving and CP-violating
processes. The agreement between the theoretical and experimental
results has shown that the CKM matrix correctly describes
these measurements of $\beta$ to good precision.

A major goal of the \BF{} experiments is now to search for indirect
evidence of New Physics (NP).  One strategy is to compare the
measured value of the CP violation (CPV) parameters
from $b \to sc\cbar$ to independent determinations of the same
quantities using processes that are sensitive to the contributions of
NP effects through loop diagrams.

CPV in B decays to a final state $f$ can be parameterized by \cf{},
measuring direct CPV, and \sf, measuring CPV in the interference
between decays with and without mixing. In the Standard Model for
penguin-dominated processes $b \to s\q\qbar$
$(\q=u,d,s)$~\footnote{Unless explicitly stated, conjugate reactions
  are assumed throughout this paper.}, \sf{} and \cf{} are expected to
be consistent with the values from $b \to \s\c\cbar$
decays. Additional CKM suppressed contributions to the amplitude can
induce only small deviations from this expectation. On the other hand,
additional loop contributions from NP processes may produce observable
deviations~\cite{Grossman:1996ke,Ciuchini:1997zp}.

The CKM and color suppression of the tree-level $b\to \s\u\ubar$
transition leads to the expectation that the decay \Bztokspiz{} is
dominated by a top quark mediated $\b\to\s\d\dbar$ penguin diagram,
which carries a weak phase $\arg(V_{\t\b}V_{\t\s}^*)$. If non-leading
contributions are small, \skspiz{} is expected to be equal to
\sintwobeta{} and $\ckspiz\simeq 0$.

In addition, it is possible to combine the direct CP asymmetries and
the branching fractions of the four $B\ra K\pi$ modes to test precise
sum rules~\cite{Deshpande:1994ii,Gronau:1995qd,Gronau:2006xu}. The
experimental uncertainty on these sum rules is dominated by the error
on the direct CP asymmetry in \Bztokzpiz{}. Therefore a precise
measurement of both the direct CP asymmetry, and the branching
fraction, in this decay channel represents an important consistency
test of the SM.

The time-dependent CP asymmetries of the decay
\Bztokspiz{}($\KS\to\pi^+\pi^-$) have been measured by
\babar~\cite{kspi0prl04} and subsequently by
Belle~\cite{kspi0belleacp}, and both experiments have also measured the
branching ratio~\cite{kspi0bellebf,kspi0prd05}. In this work, we
present an update of the these results based on $383$ million
$\Y4S\to\BB$ decays collected with the \babar\ detector at the PEP-II
$\epem$ collider, located at the Stanford Linear Accelerator Center.

The \babar\ detector, which is described elsewhere~\cite{babarDet},
provides charged particle tracking through a combination of a
five-layer double-sided silicon micro-strip detector (SVT) and a
40-layer central drift chamber, both operating in a \unit[1.5]{T}
magnetic field to provide momentum measurements. Charged kaon and pion
identification is achieved through measurements of particle energy
loss in the tracking system and Cherenkov cone angle in a detector of
internally reflected Cherenkov light.  A segmented CsI(Tl)
electromagnetic calorimeter (EMC) provides photon detection and
electron identification.  Finally, the instrumented flux return of the
magnet allows discrimination between muons and pions.

We reconstruct $\KS\to\pip\pim$ candidates from pairs of oppositely
charged tracks. The two-track combinations must form a vertex with a
$\chi^2$ probability greater than $0.001$ and a $\pip\pim$ invariant
mass within \unit[$11.2$]{\mevcc} ($3.7\sigma$) of the \KS\
mass~\cite{pdg}.  We form $\piz\to\gamma\gamma$ candidates from pairs
of energy depositions in the EMC that are isolated from any charged
tracks, carry a minimum energy of \unit[50]{\mev} per photon, fall
within the mass window \unit[$110<m_{\gamma\gamma}<160$]{\mevcc}, and
have the expected lateral shower shapes.  Finally, we construct
\Bztokspiz{} candidates by combining \KS{} and \piz{} candidates in
the event using kinematic and geometric information of the decay which
constraints the \Bz{} decay vertex to originate in the $\epem$
interaction region. We extract the flight length of the $\KS$ from the
fit and require that the reconstructed proper lifetime be greater
than five times its uncertainty. We require that the $\chi^2$
probability of the fit be greater than 0.001.

For each \Bz{} candidate two, independent kinematic variables are
computed.  The first one is $\mb$, the invariant mass of the
reconstructed $B$ meson, \Brec{}.  The second one is $\mmiss$, the
invariant mass of the other $B$, \Btag{}, computed from the known beam
energy, applying a mass constraint to \Brec{}~\cite{kspi0prd05}.  For
signal decays, $\mb$ ($\mmiss$) peaks near the \Bz{} mass with a
resolution of \unit{$\sim 36$}{\mevcc} (\unit[$\sim5.3$]{\mevcc}).
Both the $\mmiss$ and $\mb$ distributions exhibit a low-side tail from
leakage of energy deposits out of the EMC.  We select candidates
within the window \unit[$5.11<\mmiss<5.31$]{\gevcc} and
\unit[$5.13<\mb<5.43$]{\gevcc}, which includes the signal peak and a
``sideband'' region for background characterization. For
the~\unit[$0.8$]{\%} of events with more than one reconstructed
candidate, we select the combination with the smallest
$\chi^2=\sum_{i=\piz,\KS} (m_i-m'_i)^2/\sigma^2_{m_i}$, where $m_i$
($m'_i$) is the measured (nominal) mass and $\sigma_{m_i}$ is the
estimated uncertainty on the measured mass of particle $i$.

We exploit topological observables, computed in the $\Upsilon(4S)$
rest frame, to discriminate jet-like $\epem$ to $\qqbar$ events
$(q=u,d,s,c)$ from more spherical \BB{} events.  We compute the value
of $L_2/L_0$, where $L_j\equiv\sum_i |{\bf p}^*_i| |\cos
\theta^*_i|^j$.  Here, ${\bf p}^*_i$ is the momentum of particle $i$
and $\theta^*_i$ is the angle between ${\bf p}^*_i$ and the sphericity
axis~\cite{Bjorken:1969wi} of the \Brec{} candidate, and the sum does
not include the decay tree of the \Btag.  In order to reduce the
number of background events, we require $L_2/L_0<0.55$. We compute
$\costhetacms$, the cosine of the angle between the direction of the
$B$ meson and the nominal direction of the magnetic field ($z$ axis).
This variable is distributed as $1-\cos^2\thetacms$ for signal events
and nearly flat for background events. We select events with
$|\costhetacms| < 0.9$. We also use the distribution of $L_2/L_0$ and
of $\costhetacms$ to discriminate the signal from the residual
background in a maximum likelihood fit.  Using a full detector
simulation, we estimate that our selection retains $(33.6 \pm 1.6)\%$
of the signal events, where this error includes statistical and
systematic contributions.  The selected sample of \Bztokspiz{}
candidates is dominated by random $\KS\piz$ combinations from
$\epem\to\qqbar$ $(\q=u,d,s,c)$ fragmentation.  Using large samples of
simulated \BB{} events, we find that backgrounds from other \B{} meson
decays can be generally neglected, but we include some specific $B$
decay channels in our study of the systematic errors.

For each \Bztokspiz{} candidate, we examine the remaining tracks and
neutral candidates in the event to determine if the \Btag{} meson
decayed as a \Bz{} or a \Bzb{} (flavor tag).  We use a neural network
to determine the flavor of the $\Btag$ meson from kinematic and
particle identification information~\cite{ref:sin2betaPRL04}. Each
event is assigned to one of six mutually exclusive tagging
categories, designed to combine flavor tags with similar performance
and vertex resolution.  We measure the performance of this
algorithm in a data sample ($B_{\rm flav}$) of fully reconstructed
$\Bz\to D^{(*)-} \pip/\rho^+/a_1^+$ decays. The average effective
tagging efficiency obtained from this sample is $Q = \sum_c
\epsilon_S^c (1-2w^c)^2=(30.5 \pm 0.3)\%$, where $\epsilon_S^c$ and
$w^c$ are the signal efficiency and mistag probability,
respectively, for events tagged in category $c$, and the error is
statistical only. We take into account differences in tagging
efficiency (for signal and background) and mistag (only for signal)
for $\Bz$ and $\Bzb$ events, in order to exclude any source of fake
CPV effects.  For the background, the fraction of events
($\epsilon_B^c$) and the asymmetry in the rate of $\Bz$ versus $\Bzb$
tags in each tagging category are extracted from the fit to the 
$B_{\rm flav}$ data.

Time-dependent CP asymmetries are determined from the
distribution of the difference of the proper decay times,
$\deltat\equiv t_{\CP}-t_\text{tag}$, where the $t_{\CP}$ refers to
the \Brec{} and $t_\text{tag}$ to the \Btag{}. At the
$\Upsilon(4S)$ resonance, the $\deltat$ distribution follows
\begin{eqnarray}
  \label{eqn:td} 
\lefteqn{
{\cal P}(\deltat) \;  = 
\; \frac{e^{-|\deltat|/\tau}}{4\tau}
} &&  \nonumber \\ 
& \times \{\: 1 \; \pm \; \left[\: 
     S_f \sin{( \deltamd\deltat)} - 
     C_f \cos{( \deltamd\deltat)} \: 
     \right]\: \}\;, &
\end{eqnarray}
where the +(-) sign corresponds to \Btag{} decaying as \Bz{} (\Bzb),
$\tau$ is the neutral \B{} lifetime, \deltamd{} is the mixing angular
frequency, $C_f$ is the magnitude of direct CP violation in the decay
to final state $f$, and $S_f$ is the magnitude of CP violation in the
interference between mixing and decay. To account for flavor mistags
we reduce \sf{} by the factor $1-2w^c$. For the case of penguin
dominance, we expect $S_{\KS\piz}\simeq\sin2\beta$, and
$C_{\KS\piz}\simeq 0$.

The reconstructed proper time difference \deltatrec{} is computed from
the measured $\deltaz=\zrec-\ztag$, the difference of the
reconstructed decay vertex positions of the \Brec{} and \Btag{}
candidates along the boost direction, and the known boost of the
\epem{} system.  A description of the inclusive reconstruction of the
\Btag{} vertex is given in \cite{ref:Sin2betaPRD}.  For the
\Bztokspiz{} decay, where no charged particles are present at the
decay vertex, we identify the vertex of the \Brec{} using the single
\KS{} trajectory from the $\pi^+ \pi^-$ momenta and the knowledge of
the average interaction point (IP)~\cite{kspi0prl04}, which is
determined several times per hour from the spatial distribution of
vertices from two-track events.  We compute \deltatrec{} and its
uncertainty from a geometric fit to the $\Upsilon(4S)\to\Bz\Bzb$
system that takes this IP constraint into account. We further improve
the sensitivity to \deltatrec{} by constraining the sum of the two $B$
decay times ($t_{\CP}+t_\text{tag}$) to be equal to $2\:\tau$ with an
uncertainty $\sqrt{2}\; \tau$, which effectively constrains the two
vertices to be near the \Y4S{} line of flight. We have verified in a
full detector simulation that this procedure provides an unbiased
estimate of \deltat{}.

The per-event estimate of the uncertainty on \deltatrec{} reflects the
strong dependence of the \deltat{} resolution on the $\KS$ flight
direction and on the number of SVT layers traversed by the $\KS$ decay
daughters. In about \unit[$60$]{\%} of the selected events, each pion
track is reconstructed from at least one $\phi$ hit and one $z$ hit in
the first three layers, leading to a sufficient resolution for the
time-dependent CPV measurement. The average \deltat{} resolution in
these events is about \unit[$1.0$]{ps}. For events which fail this
criterion or for which \unit[$\deltatrec>20$]{ps} or the error on 
\deltatrec{} satisfies \unit[$\sigma_{\deltatrec}>2.5$]{ps}, the
\deltatrec{} information is not used.  However, since \cf{} can also be
extracted from flavor tagging information alone, these events still
contribute to the measurement of \cf{} and to the signal yield.

We obtain the probability density function (PDF) for the
time-dependence of signal decays from the convolution of
Eq.~\ref{eqn:td} with a resolution function ${\cal R}(\delta t \equiv
\deltatrec -\deltat,\sigma_{\deltatrec})$, where $\deltat$ is the true
value of the proper time difference from Monte Carlo. The resolution
function is parameterized as the sum of a core and a tail Gaussian,
each with a width proportional to the reconstructed
$\sigma_{\deltatrec}$, and a third Gaussian centered at zero with a fixed
width of \unit[$8$]{ps}~\cite{ref:Sin2betaPRD}.  We have verified in
simulation that the parameters of ${\cal R}(\delta t,
\sigma_{\deltatrec})$ for \Bztokspiz\ decays are similar to those
obtained from the $B_{\rm flav}$ sample, even though the distributions
of $\sigma_{\deltatrec}$ differ considerably. Therefore, we extract these
parameters from a fit to the $B_{\rm flav}$ sample.  We find that the
\deltatrec{} distribution of background candidates is well described
by a $\delta$ function convolved with a resolution function with the
same functional form as used for signal events. The parameters of the
background function are determined together with the CPV parameters
and the signal yield.

We extract the CPV parameters from an extended unbinned
maximum-likelihood (ML) fit to kinematic, event shape, flavor tag, and
decay time variables.  We have verified that all correlations
are negligible, so we construct the likelihood from the product of
one-dimensional PDFs.  Residual correlations are taken into account in
the systematic uncertainty, as explained below.

The PDFs for signal events are parameterized from a large sample of
fully-reconstructed $B$ decays in data and from simulated events.  For
background PDFs, we select the functional form from from the
background-dominated sideband regions in our data.

The likelihood function is defined as:
{\small\begin{eqnarray}
\label{eq:ml}
\lefteqn{
{\cal L}(\sf,\cf,N_S,N_B,f_S,f_B,\vec{\alpha}) =\frac{e^{-(N_S+N_B)}}{N\,!}} &&  \\
&& \times \prod_{i \in g}
    \left[ N_S f_S \epsilon^{c}_S{\cal P}_S(\vec{x}_i,\vec{y}_i;\sf,\cf) + 
     N_B f_B \epsilon^{c}_B {\cal P}_B(\vec{x}_i,\vec{y}_i;\vec{\alpha}) \right]
  \nonumber\\ 
&& \times \prod_{i \in  b}
    \left[ N_S (1-f_S) \epsilon^{c}_S {\cal P}'_S(\vec{x}_i;\cf) + 
     N_B (1-f_B) \epsilon_B^{c} {\cal P}'_B(\vec{x}_i;\vec{\alpha}) \right], \nonumber
\end{eqnarray}} \noindent
where the $N$ selected events are partitioned into two subsets: $i \in
 g$ events have $\deltatrec$ information, while $i \in b$ events do
 not. $f_S$ ($f_B$) is the fraction of signal (background) events $\in
 g$, and the complement to one is the fraction of events $\in b$. The
 probabilities ${\cal P}_S$ and ${\cal P}_B$ are products of PDFs for
 signal ($S$) and background ($B$) hypotheses evaluated for the
 measurements
 $\vec{x}_i=\{\mb,\;\mmiss,\;L_{2}/L_{0},\;\costhetacms,\;\text{flavor
 tag},\;\text{tagging category}\}$ and
 $\vec{y}_i=\{\deltatrec,\sigma_{\deltatrec}\}$.  ${\cal P}'_S$ and
 ${\cal P}'_B$ are the corresponding probabilities for events without
 $\deltatrec$ information.  In the formula, $\vec{\alpha}$ represents
 the set of parameters that define the shape of the PDFs.  Along with
 the CP asymmetries \sf\ and \cf, the fit extracts the yields $N_S$
 and $N_B$, the fraction of events $f_S$ and $f_B$, and the parameters
 $\vec{\alpha}$ which describe the background PDFs.

Fitting the data sample of $18111$ $\Bztokspiz$ candidates, we find
\mbox{$N_S=459 \pm 29$} signal decays with \mbox{$\skspiz = 0.40 \pm
0.23$} and \mbox{$\ckspiz = 0.24 \pm 0.15$}, where the uncertainties
are statistical only. The linear correlation coefficient between the
CPV parameters is $-7.3\%$.  Taking into account the selection
efficiency and the total number of \BB{} pairs in the data sample, we
obtain the branching fraction ${\cal B}(\Bztokspiz)= (10.4 \pm
0.7)\times 10^{-6}$ which does not include systematic corrections
on the yield.

\begin{figure}[!tbp]
\begin{center}
\includegraphics[width=1\linewidth]{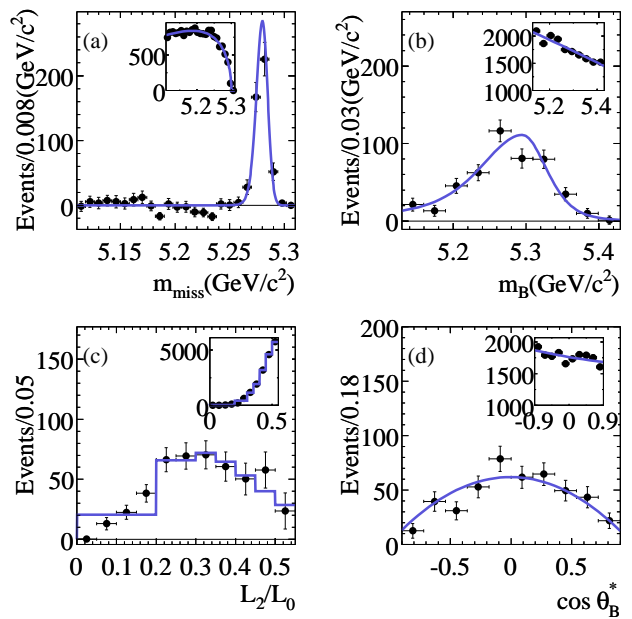}
\caption{Distribution of (a) $\mmiss$, (b) $\mb$, (c) $L_2/L_0$,
(d) $\costhetacms$, for background subtracted events on data
(dots).  The solid curve represents the shape of signal PDF, as
obtained from the ML fit. The insets show the distribution of the data,
and the PDF, for signal subtracted events. The binning of the $L_2/L_0$
PDF is coarser where the signal is well separated from the background  to
reduce the number of free parameters.}
\label{fig:splots}
\end{center}
\end{figure}

\begin{figure}[!tbp]
\begin{center}
\includegraphics[width=0.9\linewidth]{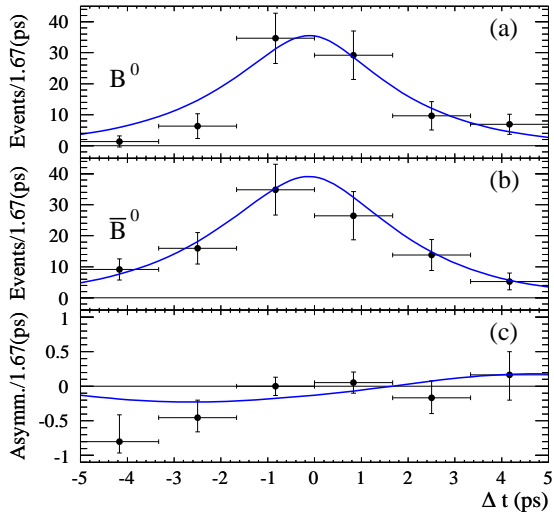}
\end{center}
\caption{Background subtracted distributions of $\deltatrec$ for events
  where $B_{\rm tag}$ is tagged as (a) $\Bz$ or (b) $\Bzb$, and (c)
  the asymmetry ${\cal A}(\deltatrec)$.  The points are data and the
  curves are the PDF projections. }
\label{fig:dtplot}
\end{figure}

Fig.~\ref{fig:splots} shows distributions for signal (background)
events, where background (signal) is subtracted using an event
weighting technique~\cite{splot}.  Figure~\ref{fig:dtplot} shows
distributions of $\deltatrec$ for $\Bz$ and $\Bzb$-tagged events, and
the asymmetry ${\cal A}_{\KS\piz}(\deltatrec) = \left[N_{\Bz} -
N_{\Bzb}\right]/\left[N_{\Bz} + N_{\Bzb}\right]$ as a function of
$\deltatrec$, for background subtracted events.  $N_{\Bz}$
($N_{\Bzb}$) represents the number of events tagged as $\Bz$ ($\Bzb$).

In order to validate the IP-constrained vertexing technique for CPV
measurements we examine \Bztojpsiks{} decays in data, where
$\jpsi\to\mup\mun$ or $\jpsi\to \epem$.  In these events we determine
$\deltatrec$ in two ways: by fully reconstructing the $\Bz$ decay
vertex using the trajectories of charged daughters of the $\jpsi$ and
the $\KS$ mesons (standard method), or by neglecting the $\jpsi$
contribution to the decay vertex and using the IP constraint and the
\KS{} trajectory only. This study shows that within statistical
uncertainties, the IP-constrained $\deltatrec$ measurement is unbiased
with respect to the standard technique and that the fit values of
$S_{\jpsi\KS}$ and $C_{\jpsi\KS}$ are consistent between the two
methods.

To compute the systematic error associated with the signal yield and
CPV parameters, each of the input parameters to the likelihood fit is
shifted by $\pm 1\sigma$ from its nominal value and the fit is
repeated.  Here, $\pm 1 \sigma$ is the associated error, as obtained
from the \Bflav{} sample (for $\Delta t$ and tagging) or from Monte
Carlo.  This contribution to the systematic error takes into account
the limited statistics we used to parametrize the shape of the
likelihood in Eq.~\ref{eq:ml}.  We find a systematic error of $0.72$
events on the yield, and of $0.006$ ($0.010$) on \skspiz{}
(\ckspiz{}).  As an additional systematic error associated with the
shape of the PDF, we also quote the largest deviation observed when
the parameters of the individual signal PDFs are floated in the
fit. This gives a systematic error of $11$ events on the yield, and of
$0.019$ ($0.018$) on \skspiz{} (\ckspiz{}). The output values of the
PDF parameters are also used to assign a systematic error to the
selection efficiency of the cuts on the likelihood variables.
Comparing the efficiency to the Monte Carlo simulation we obtain a
relative systematic error of $3.7\%$. We evaluate the systematic error
coming from the neglected correlations among fit variables using a set
of simulated Monte Carlo experiments, in which we embed signal events
from a full detector simulation with events generated from the
background PDFs. Since the shifts are small and only marginally
significant we use the average shift in yield ($-2.3$ events) and CPV
parameters ($-0.003$ on \skspiz{} and $+0.015$ on \ckspiz{}) as the
associated uncertainty.

We estimate the background from other $B$ decays to be small in the
nominal fit. We account for a systematic shift induced on the signal
yield and a small systematic uncertainty induced on the CPV parameters
by this neglected component by embedding simulated $B$ background
events in the dataset and evaluating the average shift in the fit
result: $+4.5$ events on the signal yield, $+0.003$ on \skspiz{} and
$-0.002$ on \ckspiz{}. We adjust the signal yield accordingly and we
use half of the shift as a systematic uncertainty.

To quantify possible additional systematic effects, we examine large
samples of simulated \Bztokspiz\ and \Bztojpsiks\ decays. We employ
the difference in resolution function parameters extracted from these
samples to evaluate uncertainties due to the use of the resolution
function ${\cal R}$ extracted from the $B_{\rm flav}$ sample. 
We also use the data-Monte Carlo difference of the resolution function
in \Bztojpsiks\ decays to quantify possible problems in the
reconstruction of the $\KS$ vertex. We obtain a combined systematic
error from this control sample of $0.027$ on \skspiz{} and $0.006$ on
\ckspiz{}.

We assign a systematic uncertainty of $0.002$ on \skspiz{} and $0.001$
on \ckspiz{} to account for possible misalignments of the SVT. This
does not include the effects associated with changes in the \deltat
resolution function since this is measured on data using the $B_{\rm
flav}$ control sample. We consider large variations of the IP position
and resolution, which produce a systematic uncertainty of $0.004$ on
\skspiz{} and $0.001$ on \ckspiz{}.  Additional contributions come
from the error on the known $B^0$ lifetime ($0.0022$ on both \skspiz{}
and \ckspiz{}), the value of $\Delta m_d$ ($0.0017$ on both \skspiz{}
and \ckspiz{}), and the effect of interference on the tag
side~\cite{interference} ($0.0014$ on \skspiz{} and $0.014$ on
\ckspiz{}). The systematic uncertainties on \skspiz{} and \ckspiz{}
are summarized in Table~\ref{tab:SCsys}.

For the branching fraction, systematic errors come from the knowledge
of selection efficiency, $(33.6 \pm 1.6)\%$, the counting of \BB{}
pairs in the data sample, $(383.2 \pm 4.2)\times 10^{6}$ $\BB$ pairs,
and the branching fractions of the $B$ decay chain ${\cal B}(K^0_S
\to \pi^+\pi^-)=0.6920 \pm 0.0005$ and ${\cal B}(\pi^0 \to \gamma
\gamma) = 0.9880 \pm 0.0003$)~\cite{pdg}. The systematic uncertainties on
the BF are summarized in Table~\ref{tab:BFsys}.

\begin{table}[htb]
  \begin{center}
    \caption{Summary of contributions to the systematic error on \skspiz
      and \ckspiz\label{tab:SCsys}.}
    \begin{tabular}{lcc}
      \hline\hline
      & $\Delta S(10^{-2})$ & $\Delta C(10^{-2})$ \\
      \hline
      Stat. precision of PDF parameters          & $0.6$    &  $1.0$ \\
      Shape of signal PDF                        & $1.9$    &  $1.8$\\ 
      $\BB$ background                           & $0.3$    &  $0.2$ \\ 
      Correlation among fit observables          & $0.3$    &  $1.5$ \\ 
      Vertexing method and ${\cal R}(\deltatrec, \sigma_{\deltatrec})$ & $2.7$    &  $0.6$ \\
      SVT alignment                              & $0.2$    &  $0.1$ \\
      Beam spot position calibration             & $0.4$    &  $0.1$ \\
      \Bz{} lifetime                             & $0.2$    &  $0.2$ \\ 
      Mixing frequency                           & $0.2$    &  $0.2$ \\
      Tag side interference                      & $0.1$    &  $1.4$\\
      \hline
      Total                                      & $3.4$    &  $3.0$ \\
      \hline\hline
    \end{tabular}
  \end{center}
\end{table}

\begin{table}[htb]
  \begin{center}
    \caption{Summary of dominant contributions to the systematic error
    on the branching fraction\label{tab:BFsys}.}
    \begin{tabular}{lc}
      \hline\hline              & $\Delta {\cal B} / {\cal B} (\%)$  \\
      \hline
      $\pi^0$ efficiency                         & $3.0$ \\
      $\KS$ efficiency                           & $0.6$ \\
      Cut on likelihood variables                & $3.7$ \\
      Shape of signal PDF                        & $2.5$ \\ 
      $\BB$ background                           & $0.5$  \\ 
      Correlation among fit observables          & $0.5$  \\ 
      Number of $\BB$                            & $1.1$  \\
      Resolution function                        & $0.2$  \\
      \hline
      Total                                      & $5.6$  \\
      \hline\hline
    \end{tabular}
  \end{center}
\end{table}

In summary, we have performed a measurement of the time-dependent CP
asymmetries of \Bztokspiz{} and the branching fraction of \Bztokzpiz.
We measured the CPV parameters $\ckspiz = 0.24 \pm 0.15 \pm 0.03$ and
$\skspiz = 0.40 \pm 0.23 \pm 0.03$, and the branching fraction ${\cal
  B}(\Bztokzpiz)= (10.3 \pm 0.7 \pm 0.6)\times 10^{-6}$.  The first
errors are statistical and the second ones are systematic.  These
values are consistent with the Standard Model predictions and the
experimental value of $\sin2\beta$. The results presented in this work
supersede previous ones~\cite{kspi0prd05}. Using the rate sum rule
from~\cite{Gronau:2006xu} and the currently published results for the
other three $B\ra K\pi$ modes we find a prediction for
$\BR(\Bztokspiz)^{sr}= (9.0 \pm 0.7)\times 10^{-6}$ which is
consistent with our experimental result. Using this result the
difference between the experimental result and the prediction improves
from $1.3 \pm 1.1$ to $0.9 \pm 1.0$, which is consistent with zero.

\par
We are grateful for the excellent luminosity and machine conditions
provided by our \pep2\ colleagues, 
and for the substantial dedicated effort from
the computing organizations that support \babar.
The collaborating institutions wish to thank 
SLAC for its support and kind hospitality. 
This work is supported by
DOE
and NSF (USA),
NSERC (Canada),
CEA and
CNRS-IN2P3
(France),
BMBF and DFG
(Germany),
INFN (Italy),
FOM (The Netherlands),
NFR (Norway),
MIST (Russia),
MEC (Spain), and
STFC (United Kingdom). 
Individuals have received support from the
Marie Curie EIF (European Union) and
the A.~P.~Sloan Foundation.

\end{document}